\documentclass[twocolumn]{aastex62}
\usepackage{amsmath}

\newcommand{\cm}{{\rm cm}}

\newcommand{\gr}{{\rm g}}

\newcommand{\ye}{\ensuremath{ Y_{\rm e}}}
\newcommand{\Cf}{{$^{254}$Cf}}

\graphicspath{{./}{figs/}}

\shorttitle{Californium-254 and kilonova light curves}
\shortauthors{Y. Zhu et al.}
\begin{document}

%%%%%%%%%%%%%%%%%%%%%%%%%%%%%%%%%%%%%%%%%%%%%%%%%%%%%%%%%%%%%%%%%%%%%%%%%%%%%%

\title{Californium-254 and kilonova light curves}

%%%%%%%%%%%%%%%%%%%%%%%%%%%%%%%%%%%%%%%%%%%%%%%%%%%%%%%%%%%%%%%%%%%%%%%%%%%%%%

% the following line is for submission, including submission to the arXiv!!
\hspace{5.2in} \mbox{LA-UR-18-25559}

\correspondingauthor{M.~R. Mumpower}
\email{mumpower@lanl.gov}

%%%%%%%%%%%%%%%%%%%%%%%%%%%%%%%%%%%%%%%%%%%%%%%%%%%%%%%%%%%%%%%%%%%%%%%%%%%%%%

\author[0000-0003-0245-827X]{Y. Zhu}
\affiliation{North Carolina State University, Raleigh, NC 27695-8202, USA}

\author[0000-0003-3265-4079]{R.~T. Wollaeger}
\affiliation{Center for Theoretical Astrophysics, Los Alamos National Laboratory, Los Alamos, NM, 87545, USA}

\author[0000-0002-3305-4326]{N. Vassh}
\affiliation{University of Notre Dame, Notre Dame, Indiana 46556, USA}

\author[0000-0002-4729-8823]{R. Surman}
\affiliation{University of Notre Dame, Notre Dame, Indiana 46556, USA}
\affiliation{Joint Institute for Nuclear Astrophysics - Center for the Evolution of the Elements, USA}

\author[0000-0002-4375-4369]{T.~M. Sprouse}
\affiliation{University of Notre Dame, Notre Dame, Indiana 46556, USA}

\author[0000-0002-9950-9688]{M.~R. Mumpower}
\affiliation{Center for Theoretical Astrophysics, Los Alamos National Laboratory, Los Alamos, NM, 87545, USA}
\affiliation{Joint Institute for Nuclear Astrophysics - Center for the Evolution of the Elements, USA}
\affiliation{Theoretical Division, Los Alamos National Laboratory, Los Alamos, NM, 87545, USA}

\author[0000-0002-5848-3565]{P. M{\"o}ller}
\affiliation{Joint Institute for Nuclear Astrophysics - Center for the Evolution of the Elements, USA}
\affiliation{Theoretical Division, Los Alamos National Laboratory, Los Alamos, NM, 87545, USA}

\author[0000-0001-6811-6657]{G.~C. McLaughlin}
\affiliation{North Carolina State University, Raleigh, NC 27695-8202, USA}
\affiliation{Joint Institute for Nuclear Astrophysics - Center for the Evolution of the Elements, USA}

\author[0000-0003-4156-5342]{O. Korobkin}
\affiliation{Center for Theoretical Astrophysics, Los Alamos National Laboratory, Los Alamos, NM, 87545, USA}
\affiliation{Joint Institute for Nuclear Astrophysics - Center for the Evolution of the Elements, USA}

\author[0000-0001-7463-4899]{T. Kawano}
\affiliation{Theoretical Division, Los Alamos National Laboratory, Los Alamos, NM, 87545, USA}

\author[0000-0002-2929-1740]{P.~J. Jaffke}
\affiliation{Theoretical Division, Los Alamos National Laboratory, Los Alamos, NM, 87545, USA}

\author[0000-0002-5463-6800]{E.~M. Holmbeck}
\affiliation{University of Notre Dame, Notre Dame, Indiana 46556, USA}
\affiliation{Joint Institute for Nuclear Astrophysics - Center for the Evolution of the Elements, USA}

\author[0000-0003-2624-0056]{C.~L. Fryer}
\affiliation{Center for Theoretical Astrophysics, Los Alamos National Laboratory, Los Alamos, NM, 87545, USA}
\affiliation{Joint Institute for Nuclear Astrophysics - Center for the Evolution of the Elements, USA}

\author[0000-0002-5412-3618]{W.~P. Even}
\affiliation{Center for Theoretical Astrophysics, Los Alamos National Laboratory, Los Alamos, NM, 87545, USA}
\affiliation{Joint Institute for Nuclear Astrophysics - Center for the Evolution of the Elements, USA}
\affiliation{Department of Physical Science, Southern Utah University, Cedar City, UT 84720, USA}

\author[0000-0002-0861-3616]{A.~J. Couture}
\affiliation{Center for Theoretical Astrophysics, Los Alamos National Laboratory, Los Alamos, NM, 87545, USA}
\affiliation{Joint Institute for Nuclear Astrophysics - Center for the Evolution of the Elements, USA}

\author[0000-0003-3340-4784]{J. Barnes}
\affiliation{Department of Physics and Columbia Astrophysics Laboratory, Columbia University, New York, NY 10027, USA.}
\affiliation{Einstein Fellow}

%%%%%%%%%%%%%%%%%%%%%%%%%%%%%%%%%%%%%%%%%%%%%%%%%%%%%%%%%%%%%%%%%%%%%%%%%%%%%
\begin{abstract}

Neutron star mergers offer unique conditions for the creation of the heavy elements and additionally provide a testbed for our understanding of this synthesis known as the $r$-process. We have performed dynamical nucleosynthesis calculations and identified a single isotope, \Cf, which has a particularly high impact on the brightness of electromagnetic transients associated with mergers on the order of 15 to 250 days. This is due to the anomalously long half-life of this isotope and the efficiency of fission thermalization compared to other nuclear channels. We estimate the fission fragment yield of this nucleus and outline the astrophysical conditions under which \Cf \ has the greatest impact to the light curve. Future observations in the middle-IR which are bright during this regime could indicate the production of actinide nucleosynthesis.

\end{abstract}

%% Keywords

%%%%%%%%%%%%%%%%%%%%%%%%%%%%%%%%%%%%%%%%%%%%%%%%%%%%%%%%%%%%%%%%%%%%%%%%%%%%%
\section{Introduction} \label{sec:intro}

With the first observation of two merging neutron stars (NSM) \citep{AbbottGW170817} theoretical predictions are for the first time being put to the test with multi-messenger observational constraints. To understand this event requires the combined effort and knowledge from a broad range of disciplines including nuclear physics,  atomic physics, astrophysics, and astronomy.

Mergers produce extreme conditions which are unattainable in the laboratory. As a result, observed signals provide a unique probe of nuclear physics and astrophysics. For example, we can test the idea of the production of heavy elements in mergers \citep{Lattimer+74} by comparing nucleosynthetic models to observed light curves \citep{Kasen+17, Tanvir+17}. Nuclear heating rates directly impact the brightness of optical/near-infrared (nIR) counterparts to NSMs -- kilonovae~\citep[or macronovae,][]{Metzger+17}, which are powered by radioactive decays much like supernovae are powered by the decay of $^{56}$Ni.

Before the light curves of supernovae were known to be dominated by the $^{56}$Ni decay chain, it was speculated that they might be driven by heavy, long-lived, neutron-rich $r$-process nuclei, specifically \Cf~\citep[e.g.][]{Baade+56, Fields+56}. At that time, experimental efforts established that spontaneous fission is the dominant decay mode of this nucleus with a half-life of $60.5\pm0.2$ days \citep{Phillips+63} and an $\alpha$-decay branching of $0.31\pm0.016\%$ \citep{Bemis+66}.

In contrast to supernovae, kilonovae from NSMs are thought to be powered by residual radioactivity of the $r$-process, and have a lanthanide-rich component. If a complete main $r$-process pattern is produced in mergers, actinides must be produced as well. In such case, fission may contribute substantially to nuclear heating due to its $\sim200$~MeV energy release. Fissioning nuclei that are likely to influence the light curve are those with half-lives on the order of days, which roughly corresponds to the predicted peak timescale~\citep{Li+98, Metzger+17}.

While the potential for the late-time dominance of \Cf has been noted in previous heating calculations \citep{Wanajo+14}, the effect of this experimentally established spontaneous fission process on the late-time light curve has not yet been explored. In this work we report that when experimentally known spontaneous fission decays are included in nucleosynthesis calculations for neutron-rich NSM ejecta, \Cf~and its fission daughter products are dominant contributors to the nuclear heating at $\sim 15-250$ days, greatly impacting late-time light curves.

%%%%%%%%%%%%%%%%%%%%%%%%%%%%%%%%%%%%%%%%%%%%%%%%%%%%%%%%%%%%%%%%%%%%%%%%%%%%%
\section{Nucleosynthesis and fission} \label{sec:nucleo}

\begin{figure*}[t]
  \centerline{
    \includegraphics[scale=0.75]{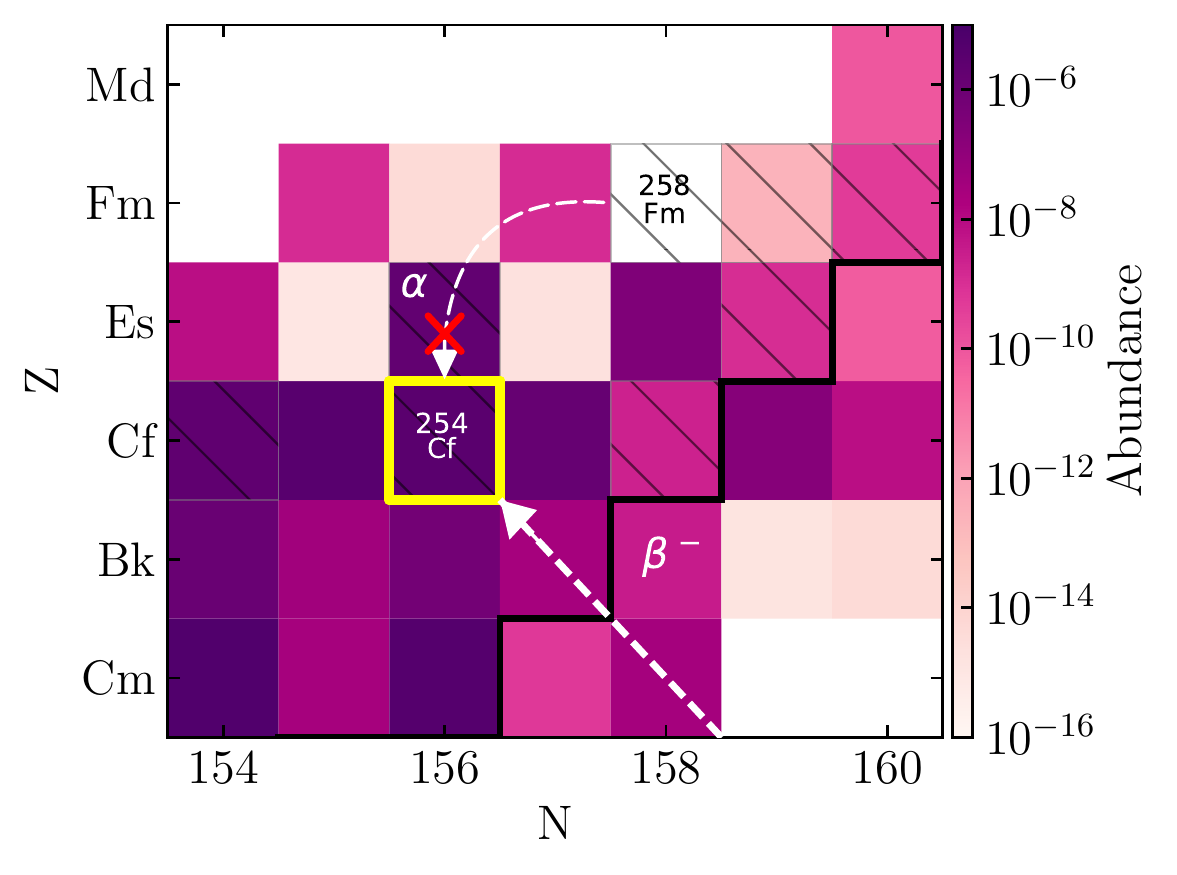}\qquad
    \includegraphics[scale=0.75]{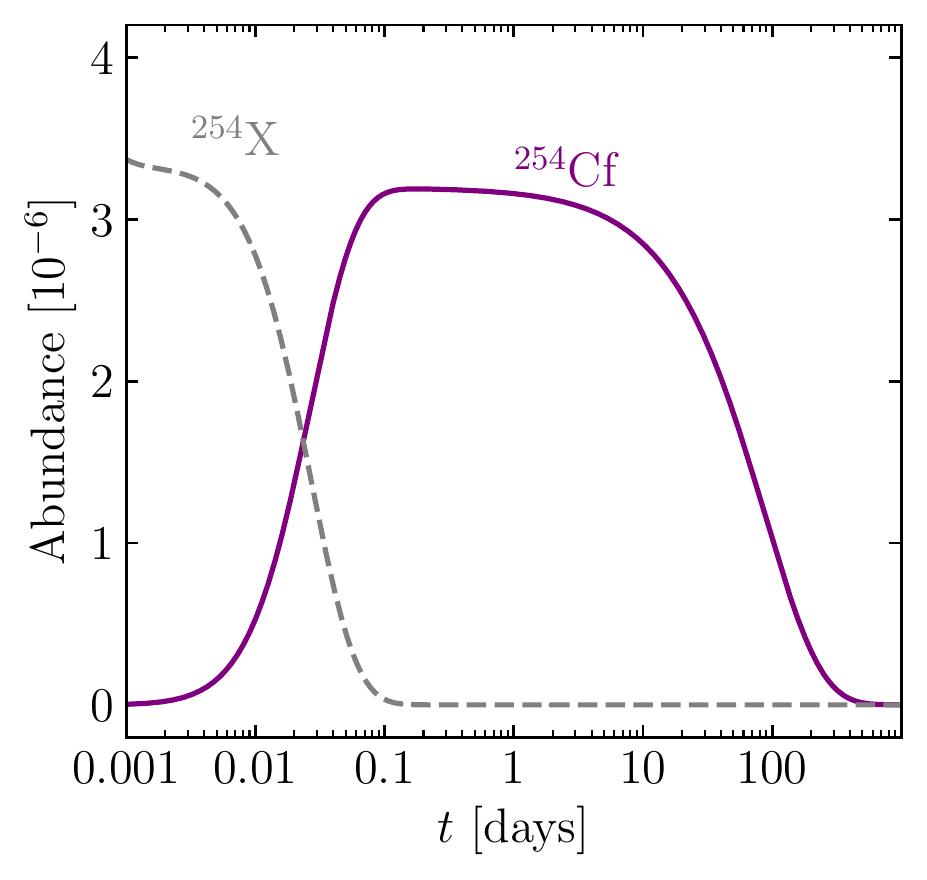}}
  \caption{\label{fig:feeding} Population of \Cf. Left: abundances of nuclei at $t=0.02$ days. The $\beta$-decay path into \Cf~and the potential $\alpha$-decay from $^{258}$Fm are shown. Nuclei that undergo spontaneous fission at this time are indicated by hatched boxes. The region to the left of the black line represents the limit of experimentally-studied nuclei. Right: the total abundance of the $A=254$ $\beta$-decay chain feeding into \Cf~over time. The \Cf~nucleus is populated only by $\beta$-decay and any possible $\alpha$-decay chains are blocked by the spontaneous fission of $^{258}$Fm.}
\end{figure*}

To model $r$-process nucleosynthesis we use the Portable Routines for Integrated nucleoSynthesis Modeling (PRISM) reaction network developed jointly at the University of Notre Dame and Los Alamos National Laboratory. In this network, now at version $2.0$, the reheating of the ejecta is handled self-consistently. We choose wind conditions that are consistent with current neutron star merger ejecta models: entropy per baryon $s/k=40$, outflow timescale $\tau=20$, and \ye~$=0.2$.

Our nucleosynthesis calculations contain all relevant nuclear reaction channels including charged particle reactions, neutron capture, photodissociation, $\beta$-decay, and delayed neutron emission. Fission from neutron-induced, $\beta$-delayed and spontaneous channels are also included. The nuclear properties are based on the theoretical nuclear model FRDM2012 \citep{Moller+15, Moller+16, Mumpower+16, Mumpower+18, Moller+18}. When available, evaluated data is used for masses \citep{AME2016} and decay properties from \texttt{NUBASE2016} \citep{NUBASE2016}. The \texttt{NUBASE2016} database contains half-lives and branching ratios for the decay channels of $\beta\pm$, electron capture, $\alpha$-decay, spontaneous fission (sf), and the spontaneous emission of neutrons and protons. For the inclusion of all such evaluated decay rates and branching ratios, care is taken to ensure no theoretical decay rates interfere with experimentally established decay processes.

The choice of theoretical sf rates produce small variation in the computed heating rates. In the calculations we present we used a parameterization found to fit within $\sim 1-3$ orders of magnitude of the measured half-lives of elements with $Z<100$ \citep{Xu+05}. Similar heating rates are obtained using phenomenological dependences on fission barrier height, such as \citep{Zagrebaev+11, Karpov+12} and \citep{Kodama+75, Petermann+12}. Nevertheless, despite recent advances \citep{Goriely+09, Giuliani+18}, the treatment of spontaneous fission in the $r$-process is subject to large uncertainties, and it is impossible to rule out unmeasured nuclei that are populated during the $r$-process and have decay timescales on the order of days.

Turning to experimentally measured data, some of the nuclei which could potentially become significantly populated in the $r$-process and undergo spontaneous fission include isotopes of Californium, Fermium, as well as $^{257}$Es and $^{260}$Md, see Fig. \ref{fig:feeding}. Although many isotopes of Fermium reportedly undergo sf, their experimentally established half-lives are on the order of seconds or faster, thus they are not populated for long enough to make significant contributions to the heating. For example, the nucleus $^{258}$Fm fissions very quickly; \texttt{NUBASE2016} reports $\sim100\%$ sf branching with a half-life of $370\pm 14$ $\mu$s. The same applies to $^{256}$Cf given its $12.3$~m half-life. This leaves $^{257}$Es, $^{260}$Md, and \Cf~  in this populated region with half-lives experimentally known to be on the order of days.

We used \texttt{NUBASE2016} branchings and so assume $^{257}$Es to undergo only $\beta$-decay, however the possibility for this nucleus to decay via sf has not been experimentally ruled out. Mendelevium-260 is known to dominantly undergo sf with half-life $\sim30$ days, however the population of $^{260}$Md is highly subject to the theoretical decay rates applied to its potential $\beta$-feeder $^{260}$Fm, which based on FRDM2012 masses has $Q_{\beta^{-}}<0$. This leaves \Cf~as a nucleus likely to influence the heating rate and light curve.

The extent of the influence of \Cf~depends on the mechanisms by which the nucleus becomes populated. Figure \ref{fig:feeding} shows that we find \Cf~to be populated solely by the $\beta$-decay of nuclei in the $A=254$ isobaric chain. As can be seen to the right of the black line in Fig. \ref{fig:feeding}, the nuclei which $\beta$-decay to populate \Cf~are for the most part unstudied. Alpha-feeding from $^{258}$Fm is prohibited by the uncertain \texttt{NUBASE2016} branching data. A non-zero $\alpha$-branching from $^{258}$Fm would seek to amplify the influence of \Cf~possibly by opening pathways to population via $\alpha$-decay chains of heavier nuclei. This highlights the importance of future experimental investigations to understand the precise branchings of nuclei in this region.

\begin{figure}
  \centerline{\includegraphics[width=85mm]{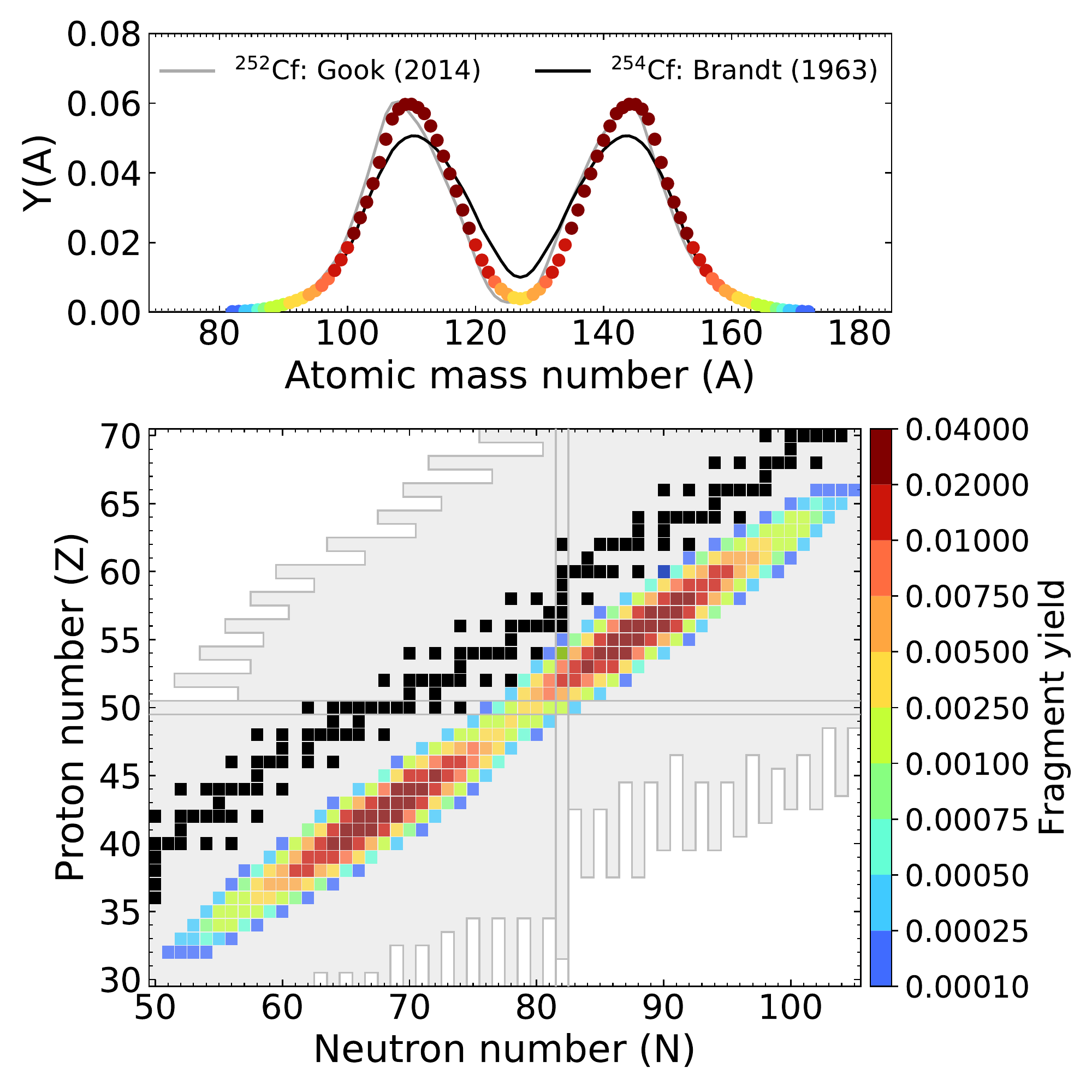}}
  \caption{Upper panel: primary fission fragment yield of \Cf(sf) calculated in the hybrid approach (see text). The experimental primary mass yield for ${}^{252}$Cf(sf) from~\cite{Gook+14} and the sparse data on \Cf(sf)~from \cite{Brandt+63} are shown for reference. Bottom panel: the two-dimensional fragment yield of \Cf(sf), with our charge distribution systematics. Stable nuclei are shaded black with the extent of FRDM2012 outlined in light gray.}
  \label{fig:fiss}
\end{figure}

We construct fission fragment yields of \Cf~with a hybrid method that combines both theoretical and experimental data. For \Cf(sf), fission fragment yields $Y(A,Z)$ in both mass $A$ and charge $Z$ are used in order to produce the most accurate estimate of the energy release. Our hybrid method is based on experimental data for the well-measured reaction $^{252}$Cf(sf) and calculations for neutron-induced fission of $^{251,253}$Cf. The available mass yields $Y(A)$ data of~\cite{Budtz+88, Hambsch+97, Zeynalov+09, Gook+14} are fit with the three-Gaussian parameterization using a global least-squares fit as done previously by~\cite{Jaffke+18}. We then calculate $Y(A)$ for the $^{251,253}$Cf($n$,f) reactions using the semi-classical method of~\cite{Randrup+11}. Next, we determine the ratio of the fitted $Y(A)$ for $^{252}$Cf(sf) over the calculated $^{251}$Cf($n$,f) at each $A$ value. This ratio is multiplied by the calculated $Y(A)$ for $^{253}$Cf($n$,f) to produce our estimate for the $Y(A)$ of \Cf(sf) shown in the top panel of Fig.~\ref{fig:fiss}.

To determine $Y(A,Z)$ we apply a charge distribution systematics $Y(Z|A)$ with $Y(A,Z) = Y(A)\times Y(Z|A)$, where

\begin{equation}
Y(Z|A) = \frac{\exp[-[Z-Z_p(A)]^2/2\sigma_Z^2]}{\sqrt{2\pi\sigma_Z^2}}
\label{eq:YZA}
\end{equation}

and the most probable charge $Z_p(A)$ is given by the unchanged charge distribution via~\cite{Wahl+88} with a charge polarization from $^{252}$Cf(sf) data of~\cite{Naik+97}. The width of the charge distribution is $\sigma_Z=0.58$. With Eq.~\ref{eq:YZA} and the hybrid $Y(A)$ for \Cf(sf), we calculate the spontaneous fission fragment yields $Y(A,Z)$ shown in the lower panel of Fig.~\ref{fig:fiss}.

%%%%%%%%%%%%%%%%%%%%%%%%%%%%%%%%%%%%%%%%%%%%%%%%%%%%%%%%%%%%%%%%%%%%%%%%%%%%%
\section{Energy Partitioning and Thermalization} \label{sec:therm}

The luminosity of a kilonova is powered by radioactivity, as the kinetic energy of the suprathermal particles emitted by nuclear decays is converted to heat. Thermal radiation then diffuses out of the ejecta cloud, producing an observable electromagnetic transient. The luminosity thus depends both on the energy released by radioactivity, and on thermalization efficiency. From a thermalization standpoint, not all decay modes are created equal. \citet{Barnes+16a} found that $\beta$-decay heats the ejecta less effectively than $\alpha$-decay, which in turn is less efficient than fission. This is partly due to the fact that $\beta$-decays release roughly 80\% of their energy in neutrinos and $\gamma$-quanta, which escape the ejecta without interacting or contributing to the heating. In contrast, in $\alpha$-decays and fission most of the energy goes to massive particles, which thermalize in the ejecta. This is compounded by the fact that fission fragments and $\alpha$-particles thermalize their energy more efficiently than $\beta$-particles via Bethe-Bloch scattering. As a result, fission and $\alpha$-decay can be extremely important sources of power for kilonovae.

To accurately account for these effects, our model of nuclear heating incorporates energy partitioning between decay products. When possible, we use  recent experimental data, provided by the Evaluated Nuclear Reaction Data Library ENDF/B-VIII.0~\footnote{\url{https://www-nds.iaea.org/public/download-endf/ENDF-B-VIII.0/}} library~\citep{Brown+18}. For nuclides not in database, we take the $Q$-value of the decay and apply the corresponding average energy partition. Given the decay rate and average energies of decay products, we calculate the total heating rate for each of the decay products, ${\{\dot\epsilon_\alpha(t),\dot\epsilon_e(t),\dot\epsilon_\gamma(t),\dot\epsilon_{\rm fis}(t)\}}$.

\begin{figure}
 \centerline{\includegraphics[width=85mm]{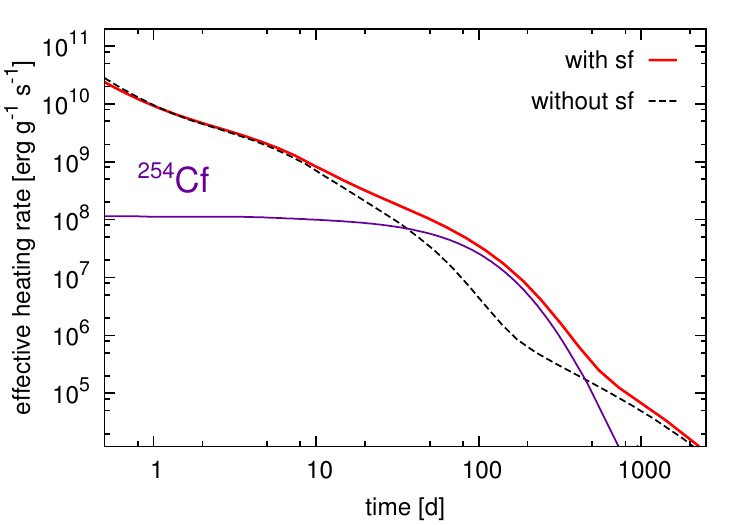}}
  \caption{Effective heating rates, including energy partitioning between decay products and their thermalization, with and without contribution from spontaneous fission of actinides, in particular \Cf. The dark purple solid line shows the (thermalized) contribution from fissioning \Cf~nuclide alone.}
  \label{fig:effheating}
\end{figure}

These quantities are combined with the analytic thermalization efficiencies calculated in \cite{Barnes+16a} to obtain the final effective heating rates~\citep[as in][]{Rosswog+17}:
\begin{align}
  \dot{\epsilon}_{\rm total}(t)
&=f_\alpha(t)\dot\epsilon_\alpha(t)
+ f_\beta(t) \dot\epsilon_e(t)
+ f_\gamma(t)\dot\epsilon_\gamma(t)
  \\
&+ f_{\rm fis}(t)\dot\epsilon_{\rm fis}(t),
\nonumber
\end{align}
where thermalization efficiencies $\{f_\alpha, f_e, f_\gamma, f_{\rm fiss}\}$ are computed as in~\cite{Barnes+16a}:
\begin{align}
  &f_\gamma(t) = 1 - \exp{\left(-\frac1{\eta_\gamma^2}\right)},
  \\
  &f_j(t) = \frac{ \log(1 + 2\eta_j^2)}{2\eta_j^2},
  \;\;\;
  j\in\{\alpha,e,{\rm fis}.\}
  \label{eq:efficiencies}
\end{align}
The dimensionless quantities $\eta_j = t/\tau_j$ are defined with respect to the thermalization timescales $\tau_j$ for each species
(${m_5\equiv m_{\rm ej}/0.05\ M_\odot}$,
 ${v_1\equiv v_{\rm ej}/0.1\ c}$):
\begin{align}
  &\tau_\gamma = 8.85\ m_5^{1/2}v_1^{-1}\;{\rm days},   \\
  &\tau_e      = 66.2\ m_5^{1/2}v_1^{-3/2}\left(\frac{0.5\ {\rm MeV}}{\langle E_e\rangle}\right)^{1/2}\;{\rm days}, \\
  &\tau_\alpha = 69.2\ m_5^{1/2}v_1^{-3/2}\left(\frac{6\ {\rm MeV}}{\langle E_\alpha\rangle}\right)^{1/2}\;{\rm days}, \\
  &\tau_{\rm fis} = 150.0\ m_5^{1/2}v_1^{-3/2}\left(\frac{125\ {\rm MeV}}{\langle E_{\rm fis}\rangle}\right)^{1/2}\;{\rm days}.
  \label{eq:timescales}
\end{align}

The final effective heating rate used for the kilonova light curve calculations is shown in Figure~\ref{fig:effheating}, for two cases: with and without inclusion of spontaneous fission. We use model values typical of those found in the literature, ${m_{\rm ej}=0.05 M_\odot}$ for the total ejecta mass, and ${v_{\rm ej}=0.1}$~c for the average velocity. Thermalization efficiencies are extremely sensitive to the expansion velocity ($\propto v_{\rm ej}^{-3}$) and to a smaller extent, mass ($\propto m_{\rm ej}$) -- for instance, doubling $v_{\rm ej}$ decreases thermalization efficiency by a factor of 8. This means that there is a degeneracy between the increased heating rate and slower or more massive ejecta. We assume that such degeneracy with respect to velocity can be resolved from observations of spectral features and with respect to mass from observations of early emission.

For the case when spontaneous heating is not included, the heating rate is dominated by $\beta$-decays~\citep{Hotokezaka+16d, Barnes+16a, Wollaeger+17}, which thermalize poorly, especially in progressively more dilute plasma. As shown in Figure~\ref{fig:effheating}, adding efficiently thermalizable spontaneous fission produces a remarkable difference, amounting to one order of magnitude higher heating at around $20$~d, and almost a factor of 100 at $100-300$~d. Perhaps even more remarkable is the fact that the difference is almost entirely due to the \Cf.

%%%%%%%%%%%%%%%%%%%%%%%%%%%%%%%%%%%%%%%%%%%%%%%%%%%%%%%%%%%%%%%%%%%%%%%%%%%%%%
\section{Light Curve Models} \label{sec:lc}

\begin{figure}
 \begin{center}
  \centerline{\includegraphics[width=85mm]{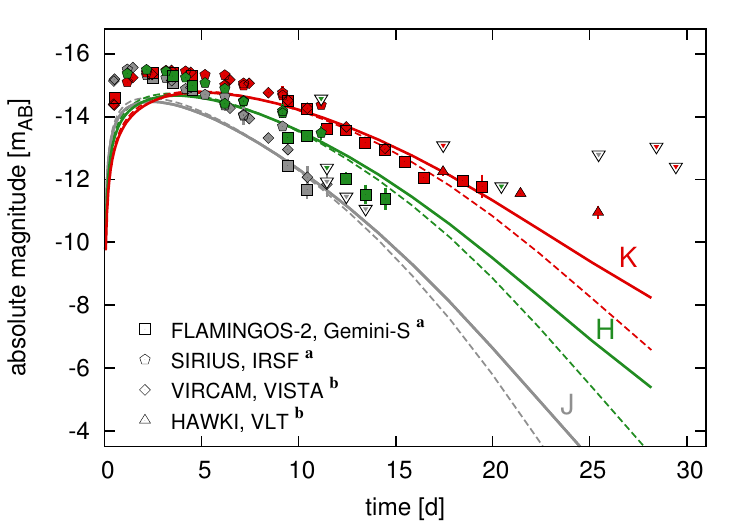}}
  \caption{Observations in the near-infrared $J$-, $H$- and $K$-bands (points),
    and the best-fit theoretical model with spontaneous fission contribution
    (solid lines). Dashed lines show theoretical light curves for the same ejecta
    mass and velocity, but without spontaneous fission contribution. At the epoch
    of 25~d, inclusion of fissioning isotope \Cf~increases the brightness
    by almost 2 mags. Observational data points are labeled by the instrument and
    the telescope which received the data. Data sources: (a)~\cite{Kasliwal+17},
    (b)~\cite{Tanvir+17}. Upturned triangles are upper limits.}
 \label{fig:lcs}
 \end{center}
\end{figure}

We use two semi-analytic light curve models to investigate whether it is possible to discern the effects of fissioning \Cf~with observations and compare our models to the near-infrared observational data from the NSM GW170817.

NSMs can produce a variety of outflows with distinct masses, velocities, and compositions~\citep{Metzger+17}. First, material torn from inspiraling neutron stars by tidal forces forms a high-velocity, low-\ye\ outflow. An additional fast, higher-\ye\ component can be produced by dynamical ``squeezing'' at the crash contact interface. Second, post-merger accretion disks may produce ``winds'' with relatively low ($0.05c - 0.1c$) velocities and a range of \ye-values, depending on weak interactions and whether the central object collapses to a black hole.  In addition a neutrino-driven wind will occur and can have relatively high \ye\ \citep{Surman+05} or, if neutrino flavor transformation is taken into account relatively low \ye ~\citep{Zhu+16}.

Low-\ye~outflows produce lanthanides and actinides with very high opacities~\citep{Kasen+13}, resulting in emission at red and near-infrared wavelengths (``red'' components). Higher-\ye\  outflows ($\ye \gtrsim 0.25$)  fail to synthesize these elements, and their emission is bluer (``blue'' components). In general, the kilonova emission will reflect contributions from both components.

An \emph{r}-process that fails to produce actinides also fails to synthesize \Cf, so heating from the fission of \Cf~is important only for low-\ye\ outflows. Since thermalization is very sensitive to mass and velocity, the effect of fission will only be prominent if the red component is slow and/or massive compared to any blue component in the model.

For example, \citet{Kasen+17} posit that the total kilonova signal is due to a fast blue outflow ejected by ``squeezing'', along with a slower, redder, lanthanide-rich disk outflow. At late times, radiation from the blue component of this model fades due to inefficient thermalization induced by rapid expansion, and the net emission is dominated by the red component. Thus, any additional effect from fission would potentially be observable in this scenario. Other models also achieved good agreement with observed broad-band light curves without including a slow red component \citep{Tanvir+17, Troja+17, Kawaguchi+18}; thus in this scenario the effect on the light curve from \Cf~fission would be minimal.

We therefore explore the effects of \Cf~fission using a single-component model with similar parameters to the red component of \citet{Kasen+17}. At the times of interest, the spectrum peaks in the near- and mid-infrared. For near-infrared emission, we use semi-analytic, spherically symmetric radiative transfer solution described in detail in~\cite{Rosswog+18} (their Appendix~A). The transfer equation is solved in a diffusion approximation, where it admits separation of variables and the solution for internal energy profiles in terms of summed radial eigenmodes with time-dependent coefficients. This method was invented by~\cite{Pinto+00} for supernova envelopes, and recently applied to kilonovae and cross-validated with the full multigroup Monte Carlo code {\tt SuperNu} \citep{Wollaeger+17}. The model employs gray opacity $\kappa=10\ \cm^2 \gr^{-1}$, which is a reasonable approximation for a lanthanide-rich ejecta \citep{Tanaka+18}.

Figure~\ref{fig:lcs} compares the resulting synthetic light curves to observational data in the near-infrared $JHK$-bands (solid lines). We employ an ejecta mass $m_{\rm ej} = 0.05\ M_{\odot}$ and a median velocity $v = 0.1\ c$. Dashed lines in the same plot show the case without spontaneous fission, resulting in light curves which are dimmer by almost two magnitudes at twenty-five days after the explosion. Here we used a \ye\ of approximately 0.2, but as the neutron richness of the ejecta is increased, the difference in the light curve due to the inclusion of \Cf~ is increased by another magnitude.

%%%%%%%%%%%%%%%%%%%%%%%%%%%%%%%%%%%%%%%%%%%%%%
\begin{figure}
 \begin{center}
  \centerline{\includegraphics[width=85mm]{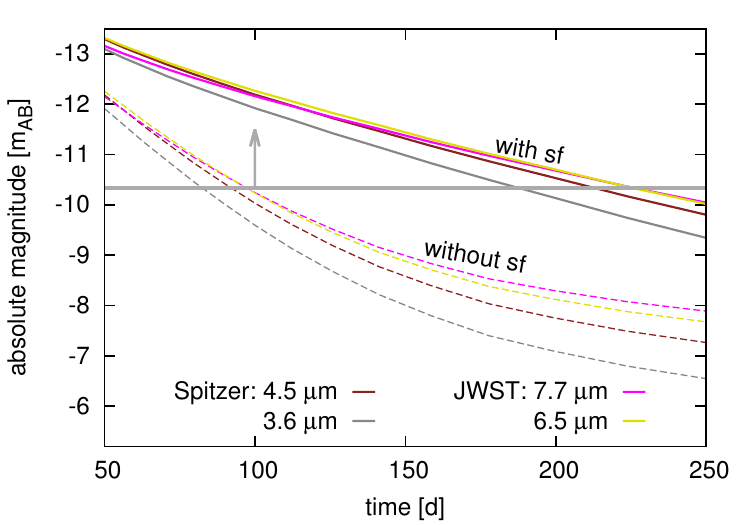}}
  \caption{Theoretical predictions for mid-IR light curves with (solid) and
     without (dashed) spontaneous fission and \Cf~contribution.
     The gray horizontal line indicates JWST sensitivity threshold for
     mergers at 200~Mpc, assuming $10$~ks exposure
     (\url{https://jwst-docs.stsci.edu/display/JTI/MIRI+Sensitivity}).}
 \label{fig:mir}
  \end{center}
\end{figure}
%%%%%%%%%%%%%%%%%%%%%%%%%%%%%%%%%%%%%%%%%%%%%%%

Figure~\ref{fig:effheating} shows that the highest impact of the heating rate from ${}^{254}$Cf is expected at late epochs, at $t\sim100-300$~d. At this time, the ejecta is optically thin, having an optical depth less than unity:
\begin{align}
\tau \approx 0.36\left(\frac{m_{\rm ej}}{0.05 M_{\odot}}
                 \cdot\frac{\kappa}{10\ {\rm cm}^2 {\rm g}^{-1}}\right)
                 \left(\frac{t}{100\ {\rm d}} \cdot \frac{v}{0.1\ c}\right)^{-2}
\end{align}

Late emission is expected to peak in the mid-infrared (IR). Because the diffusion approximation is no longer applicable in this regime, to roughly estimate the effect of \Cf~we use a separate, one-zone semi-analytic model, inspired by the original idea of \cite{Li+98}. For simplicity, we assume local thermodynamic equilibrium. The latter implies strong ion-electron coupling, which may not be satisfied at late times. However, estimating the coupling timescales \citep{Gericke+02} we find that they remain short ($< 100 {\rm s}$) compared to timescale of expansion out to 400\,d, so our assumption of strong coupling is reasonable.

In the optically thin regime, radiative internal energy of the ejecta can be estimated as:
\begin{align}
  U = \tau\cdot a T^4 V,
  \label{eq:U}
\end{align}
where $a$ is radiation density constant and V is ejecta volume. The rate of radiative energy loss is:
\begin{align}
  q_{\rm rad}=4\sigma\kappa_P\rho T^4,
  \label{eq:radloss}
\end{align}
where $\kappa_P(T)$ is the Planck mean absorption opacity.

Following the derivation in~\cite{Li+98} with expressions (\ref{eq:U}) and (\ref{eq:radloss}), it is straightforward to arrive at the following ODE (here, $\gamma\equiv 2 + c/v_{\rm ej}$ and $q_{\rm nuc}(t)$ is the effective nuclear heating rate):
\begin{align}
  \frac{d (a T^4)}{d\log{t}} = - \gamma a T^4
  + \frac{q_{\rm nuc}(t)}{\kappa_P(T) v_{\rm ej}}.
  \label{eq:aT4}
\end{align}

In the calculation below, we adopt an approximate temperature dependence of the following form:
\begin{align}
  \kappa_P(T) = {10\ \cm^2\ \gr^{-1}}\
                \left(1+\exp{\left[\dfrac{1300\ K - T}{100\ K}\right]}\right)^{-1},
   \label{eq:opac}
\end{align}
capturing an exponential drop-off in the opacity as temperature drops below 1300~K and the plasma becomes neutral~\citep[cf.][Figure 10]{Kasen+13}.

Figure~\ref{fig:mir} shows the light curves in Spitzer and JWST mid-IR bands, which come from numerically solving equation~(\ref{eq:aT4}) with opacity given by~(\ref{eq:opac}). The mass and velocity are the same as used in the previous model; the equations are integrated with the initial condition $T(10~d) = 1000$~K~\citep[cf.][]{Kasliwal+17}. A difference of almost two orders in heating translates to about four magnitudes brighter transients, which is more pronounced than in the near-IR case.

For the AT2017gfo event, no mid-IR detections were made: for Spitzer, which was the only operating mid-IR satellite in orbit at that time, the merger was outside its visibility window. However, future observations with JWST should be able to discern the presence of \Cf. As illustrated in Figure~\ref{fig:mir}, for a merger at 200~Mpc, the presence of \Cf~essentially makes a difference between detection and non-detection.

%%%%%%%%%%%%%%%%%%%%%%%%%%%%%%%%%%%%%%%%%%%%%%%%%%%%%%%%%%%%%%%%%%%%%%%%
\section{Conclusions} \label{sec:conclusions}

Unlike supernovae, electromagnetic transients associated with mergers are thought to be powered by multiple decaying nuclides. We have isolated a prominent imprint of a particular isotope -- \Cf, affecting light curves on timescales of 15-30 days in the near infrared $JHK$-bands by one to three magnitudes and at 50-250 days in the mid-infrared by almost four magnitudes. The effect of the spontaneous fission of \Cf~is most pronounced in scenarios with a significant contribution from heavy, slow outflows with low \ye. In such cases, the corresponding kilonovae should be detectable by JWST up to $250$~days.

Since \Cf~sits at a higher mass number than long-lived actinides such as $^{238}$U, the production of \Cf~implies the nucleosynthesis of at least some actinide material. Thus, a combined approach of improving experimental knowledge in this region along with the coupling of late-time light curves with nucleosynthetic simulations have the potential to play a major role in cementing the origin of the heaviest \emph{r}-process elements.

% References

\section*{Acknowledgements}

Y.Z., N.V., R.S., M.M, G.C.M., T.K. and P.J. were supported in part by the U.S. Department of Energy (DOE) under Contract No. DE-AC52-07NA27344 for the topical collaboration Fission In R-process Elements (FIRE). This work was partly supported by the U.S. DOE under Award Numbers DE-SC0013039 (R.S. and T.S.), DE-FG02-02ER41216 (Y.Z. and G.C.M.) and DE-SC0018232 (T.S.). A portion of this work was carried out under the auspices of the National Nuclear Security Administration of the U.S. DOE at Los Alamos National Laboratory (LANL) under Contract No. DE-AC52-06NA25396 (R.W., M.M., P.M., O.K., T.K., P.J., C.F., W.E., A.C.) and used resources provided by LANL Institutional Computing Program (R.W., O.K.). J.B. is supported by the National Aeronautics and Space Administration (NASA) through the Einstein Fellowship Program, grant number PF7-180162. O.K. is grateful to N. Lloyd-Ronning for valuable input. This work benefited from discussions at the 2018 Frontiers in Nuclear Astrophysics Conference supported by the National Science Foundation under Grant No. PHY-1430152 (JINA Center for the Evolution of the Elements).

\end{document}